# Improving nuclear magnetic resonance and electron spin resonance thermometry with size reduction of superparamagnetic iron oxide nanoparticles

*Pei-Yun Lin [1,2], Darshan Chalise [2,3] †, and David G. Cahill [1,2,3] ***

[1]*Materials Science and Engineering, University of Illinois at Urbana-Champaign, Urbana, IL, 61801, USA*

[2]*Materials Research Laboratory, University of Illinois at Urbana-Champaign, Urbana, IL, 61801, USA*

[3]*Department of Physics, University of Illinois at Urbana-Champaign, Urbana, IL, 61801, USA*

**Abstract**

Thermometry based on magnetic resonance has been extensively studied due to its important application in biomedical imaging. In our previous work, we showed that the spin-spin relaxation time ($T_2$) of nuclear magnetic resonance (NMR) in water is a highly sensitive thermometer as $T_2$ scales with the highly temperature-sensitive self-diffusion constant of water. In this work, in addition to temperature dependent self-diffusion constant of a fluid, we utilize the temperature dependent magnetization of 4 nm SPIONs to improve $T_2$ sensitivity ($\xi_{T_2}^T$ = 4.96) by 1.4 times over self-diffusion ($\xi_D^T$ = 3.48) alone in hexane between 248 K and 333 K. To extend the application of NMR $T_2$ thermometry to engineering systems, we also investigate the temperature dependence of $T_2$ in mineral oil (Thermo Scientific, J62592), which exhibits remarkably high sensitivity ($\xi_{T_2}^T = 11.62$) between 273 K and 353 K. This result implies that applications of NMR $T_2$ thermometry in heat transfer fluids are promising. NMR thermometry, however, is generally not applicable to solids. Therefore, we also evaluate the potential of electron spin resonance (ESR) thermometry with SPIONs in solids between 100 K and 290 K, for potential temperature monitoring in biomedical and engineering applications. The size and concentration effects on ESR signals are studied systematically, and our results show that the temperature dependent linewidth follows a $T^{-2}$ law for 4 nm SPIONs, while the concentration of SPIONs has no impact on the temperature dependence of the ESR linewidth. The linewidth at room temperature at 9.4 GHz is 10.5 mT. Combining our NMR and ESR results, we find that to obtain higher temperature sensitivity in a magnetic resonance technique using SPIONs, SPION with a small magnetic moment, i.e., a small volume and reduced magnetization, are beneficial.

* Corresponding author – d-cahill@illinois.edu

$^{†}$ Current address – Department of Materials Science, Stanford University, Stanford, CA, 94305

## I. INTRODUCTION

Several 2D thermometry techniques have been well-established to provide information on the temperature field with high sensitivity and spatial resolution [1, 2]. 3D thermometry such as conventional optical thermometry can be used for a system with optical access. For thermometry of optically opaque systems, magnetic resonance, which includes both, nuclear magnetic resonance (NMR) and electron spin resonance (ESR) provides an opportunity for sensitive 3D thermometry. Such thermometry has critical applications in biology and engineering [1-5]. For example, in medicine, NMR thermometry can be useful for hyperthermia using highly focused ultrasound (HIFU), during which heat is generated to kill tumors, and accurate monitoring of temperature is critical to protect healthy tissue [5]. In an engineering context, electronic devices are often equipped with heat transfer fluids to manage heat and therefore characterization of the temperature field in the fluid is important [6].

In our previous work in NMR thermometry, we showed, in the presence of superparamagnetic iron oxide nanoparticles (SPIONs), spin-spin relaxation time ($T_2$) is a thermometer that provides a non-invasive, highly sensitive, and high throughput 3D thermometry [3]. The spin-spin ($T_2$) relaxation of diffusing protons arises from the inhomogeneity in the magnetic field induced by SPIONs. As long as the time between the spin echo pulses ($T_{echo}$) is much longer than the correlation time ($\tau_c$) of the diffusion (where $\tau_c = d^2/4D$ for a SPION diameter $d$ and the self-diffusion constant $D$ of the proton), the spin-echo is unable to recover the relaxation and the spin-spin relaxation time ($T_2$) is solely determined by the volumetric magnetization of the SPION and the self-diffusion constant of the proton in the fluid. This differentiates $T_2$ from the relaxation $T_2^*$, resulting from the field inhomogeneity induced by the SPION as well as the inhomogeneity in the static magnetic field;

the latter of the two relaxations can be recovered using spin-echo. In our previous work [3], the temperature dependence of $T_2$ scaled only with the temperature dependence of the self-diffusion constant of the fluid as the SPIONs we used didn't show any temperature dependence of magnetization. Here, we show that the temperature dependence of $T_2$ is enhanced through the temperature dependent magnetization of small diameter (4 nm) SPIONs that have a reduced magnetization relative to larger diameter SPIONs.

A superparamagnetic nanoparticle consists of a large collection of strongly coupled spins, giving rise to a collective magnetic moment. Due to its small crystal size, thermal energy ($k_B T$) is comparable to magnetic energy, which causes thermal fluctuation of the magnetic moment, and superparamagnetic behavior is therefore exhibited. Apart from superparamagnetic behavior, it has been shown that the Curie temperature of MnFe$_2$O$_4$ nanoparticles decreases with the size due to finite size scaling [7]. As the Curie temperature decreases, it can be expected that temperature dependent magnetization of the magnetic nanoparticle increases near room temperatures. Another study also reported that iron nanoparticles with sizes under 5 nm start to exhibit significant temperature dependence of magnetization compared with larger particles [8]. From these results, it can be suggested that temperature dependence of magnetization increases as the particle size becomes smaller. In this work, we synthesize 4 nm SPIONs that show a remarkable temperature dependence of magnetization. Our results show that higher temperature sensitivity of $T_2$ is achieved in hexane and mineral oil (Thermo Scientific, J62592) with 4 nm SPIONs, combining the temperature dependence of magnetization of 4 nm SPIONs along with the temperature dependence of self-diffusion of the fluid.

Despite the advantages of NMR $T_2$ thermometry, it is limited to systems where the nucleus diffuses fast enough to give an observable NMR spectrum and is generally not applicable in most solids [4] or in liquids at low temperatures. Therefore, new techniques should be explored for such systems. Electron spin resonance (ESR) is another magnetic resonance technique which utilizes signal from magnetic moments due to unpaired electrons or free radicals [9]; and it can probe into temperature information in solids. In our previous work, we showed, ESR linewidth of a single spin in n-type semiconductors is a sensitive thermometer [4].

On the other hand, ESR linewidth of the SPIONs was reported to exhibit strong temperature dependence [10-13]. In this work, we also systematically investigate the size and concentration effects on ESR signals of SPIONs in solid polymer (tetraethylene glycol dimethacrylate, TEGDMA) and paraffin wax for best ESR thermometry between 100 K and 290 K. Thermometry in this low temperature range can be valuable for important applications, such as cryosurgery for cancer therapy [14], where real-time temperature monitor of tumor edges and ice ball margin can improve the efficiency of the treatment.

## II. EXPERIMENTAL DETAILS

Three sizes (4 nm, 4.5 nm, and 8 nm) of SPIONs were synthesized and characterized before the NMR and ESR study. Syntheses of 4.5 nm and 8 nm hydrophilic SPIONs are based on a polyol process and that of 4 nm hydrophobic SPIONs is based on a thermal decomposition method from iron-oleate. We also attempted the synthesis of 7 nm $MnFe_2O_4$ nanoparticles, due to the possible reduction in Curie temperature and subsequent improvement in temperature dependence, but it failed to show improved $T_2$ sensitivity due to poor colloidal stability in water.

### A. Synthesis

**Synthesis of 8 nm Hydrophilic SPIONs.** Our synthesis is based on a previously reported polyol method [15]. All the chemicals were used as received. 116 mg iron(III) acetylacetonate (Strem, 99%) and 8 ml triethylene glycol (Thermo Scientific, 99%) were added to a 50 ml three-necked round bottom flask equipped with a gas adaptor, condenser, and thermometer. The mixture was stirred at 500 rpm and slowly heated to 280°C under nitrogen and kept at 280°C for 30 minutes. After 30 minutes, the solution was cooled down to room temperature by removing the heat source, resulting in a black homogeneous reaction solution. The reaction solution was then transferred to four centrifugal tubes (2 ml in each tube) and 30 ml ethyl acetate was added to each tube to precipitate the black product, which was then separated by centrifugation (8000 rpm, 10 minutes) to obtain the final products.

**Synthesis of 4.5 nm Hydrophilic SPIONs.** The synthesis is based on another reported polyol method [16] where the reaction temperature and the surfactant are different from the synthesis of 8 nm SPIONs. 118 mg iron(III) acetylacetonate (Strem, 99%) and 8 ml diethylene glycol (Aldrich, 99%) were added to a 50 ml three-necked round bottom flask equipped with a gas adaptor, condenser, and thermometer. The mixture was stirred at 500 rpm, heated to 120°C under nitrogen, and kept at 120°C for 1 hour. After 1 hour, the mixture was heated to 230°C and stayed for 2 hours. After 2 hours, the reaction solution was cooled down by removing the heat source, resulting in a black homogeneous reaction solution. The reaction solution was then transferred to four centrifugal tubes (2 ml in each tube) and 30 ml ethyl acetate was added to each tube to precipitate the black product, which was then separated by centrifugation (8000 rpm, 10 minutes) to obtain the final products.

**Synthesis of 4 nm Hydrophobic SPIONs.** Synthesis consists of two steps: the preparation of the iron-oleate complex and the synthesis of 4 nm hydrophobic iron oxide nanoparticles.

The preparation of the iron-oleate complex is based on a published procedure with slight modifications [17]. All the chemicals were used as received. Firstly, 7 ml hexane, 4 ml absolute ethanol, and 3 ml nanopure water (Milli-Q) were mixed in a 50 ml round bottom flask. 1.83 g sodium oleate (TCI, 97%) and 0.54 g ferric chloride hexahydrate (Aldrich, 99%) were then added to the mixture solution. The round bottom flask was then sealed and stirred for 4 hours (500 rpm, 70°C) in a sand bath. After 4 hours, the mixture was cooled down to room temperature, and the upper oil phase was transferred to a vial (30 ml) while the bottom aqueous phase was discarded. Nanopure water (Milli-Q) was then added to the oil phase in the vial. The vial was then shaken for 15 seconds. After the oil and aqueous phase separated again, the upper washed oil phase was pipetted out to another clean vial (30 ml). This washing procedure was repeated until the bottom aqueous layer was clean, indicating that no water-soluble substance remained in the oil phase. After washing, the red-brown oil phase was transferred to a 25 ml round bottom flask and adapted to a rotary evaporator to remove excess hexane (280 rpm, 70°C). The final viscous iron-oleate complex could be obtained and stored at 4°C in the dark.

Synthesis of 4 nm hydrophobic iron oxide nanoparticles was based on previous literature [18]. The synthesis was as follows: 340 mg prepared iron-oleate complex, 0.72 ml oleyl alcohol (Thermo Scientific, 80-85%), and 1.76 ml diphenyl ether (TCI, 99%) were mixed in a 25 ml three-necked round bottom flask equipped with a gas adaptor, condenser, and thermometer. The mixture was stirred at 300 rpm under nitrogen for the reagents to mix thoroughly. After mixed thoroughly, the solution was quickly heated to 200°C (within 20 minutes) and kept at 200°C for 30 minutes. After 30 minutes, the reaction mixture was cooled down to room temperature by removing the heat source, resulting in a black-brown homogeneous colloidal suspension. The suspension was then transferred to two centrifugal tubes and 30 ml acetone was added to each tube to precipitate the products, which were then separated by centrifugation (8000 rpm, 10 minutes) to obtain the final products, and could be directly dispersed in hexane or mineral oil.

B. Material Characterization

**Transmission Electron Microscopy (TEM).** The morphology of nanoparticles was investigated with JEOL 2010 LAB6 TEM operating at 200 kV. 8 nm and 4.5 nm SPIONs are dispersed in water for imaging while 4 nm hydrophobic SPIONs are dispersed in hexane. TEM images of 8 nm, 4.5 nm, and 4 nm SPIONs are displayed in Figure 1. The particle size is analyzed with ImageJ. Due to the difficulty in drying the mineral oil, the sample of 4 nm SPIONs in mineral oil was not examined with TEM; however, the average diameter is expected to be the same as that in hexane given identical results of dynamic light scattering (DLS) and SQUID magnetometry.

**Dynamic Light Scattering (DLS).** Measurements were performed on a Malvern Zetasizer Nano ZS to examine the hydrodynamic diameter of SPIONs in hexane and mineral oil at 25°C and ensure there is no particle aggregation. A backscattering mode (173° scattering angle) was used for all measurements. Polystyrene disposable cuvettes were used for aqueous sample while a glass cuvette (Starna Cells, 1-Q-10-GL14-C) was used for hexane and mineral oil samples. For mineral oil, the dynamic viscosity was measured with a TA DHR-3 rheometer. Data on mineral oil viscosity is shown in APPENDIX A. 4 nm hydrophobic SPIONs and 8 nm hydrophilic SPIONs are used for NMR $T_2$ study and the hydrodynamic diameter in different fluids are examined with DLS. In Figure 2, the hydrodynamic diameter of 4 nm SPIONs in hexane and mineral oil are both measured to be 4.2 nm. For 8 nm SPIONs, as it is not our main subject of NMR study, the hydrodynamic diameter of 8 nm SPIONs are shown in APPENDIX B. Since 4.5 nm SPION is not used for NMR study, it was not examined with DLS.

**Superconducting Quantum Interference Device (SQUID).** Magnetic properties of SPIONs were studied with a Quantum Design MPMS 3 with a field sweep from -3 T to 3 T. Measurements on the liquid nanoparticle samples were carried out in a PTFE sample holder (Quantum Design, 8505-013). For variable temperature measurements, the temperature was regulated by PID (Proportional-Integral-Differential) control in MPMS 3. The results of SQUID measurements are displayed in Figure 3. The absence of hysteresis verifies that three types of our synthesized iron oxide nanoparticles are all superparamagnetic. Data from SQUID magnetometry is fit with the modified Langevin function [33]:

$$\mu_{SQUID} = N\mu_p \left[\coth\left(\frac{\mu_p H}{k_B T}\right) - \left(\frac{k_B T}{\mu_p H}\right)\right] + \chi H \quad (1)$$

In the equation, $\mu_{SQUID}$ is the data measured by SQUID magnetometry; $N\mu_p$ is the total magnetic moment of the sample where $N$ is the number of particles in the sample and $\mu_p$ is the magnetic moment of a particle. The liner term $\chi H$ is added to take diamagnetic background into account [19].

For NMR study, the magnetization of 4 nm and 8 nm SPION should be estimated. With average diameter of SPION determined from TEM and $\mu_p$ from SQUID, the average magnetization of the SPION can be calculated ($M_p = \mu_p/V_p$, where $V_p$ is the volume of the SPION). At room temperature, the magnetization of the 4 nm SPION is determined to be $3.3 \times 10^5$ A/m (64 emu/g, density: 5.17 g/cm³). For 8 nm SPIONs, the magnetization is $4.9 \times 10^5$ A/m (95 emu/g), which is close to the magnetization of bulk magnetite (98 emu/g) [20] due to higher reaction temperature (280°C) of polyol synthesis [15] than that of 4 nm SPIONs (200°C) from iron oleate thermal decomposition.

### C. NMR Sample and Measurements

NMR measurements were performed on a Varian Unity/Inova system using a 14.1 T (600 MHz) field and a 5-mm broadband probe. The same type of NMR tubes (Norell, XR-55-7) was used for all measurements. Temperature was calibrated with temperature dependent chemical shift difference between $CH_2$ and hydroxyl group of ethylene glycol above 35°C and between $CH_3$ and hydroxyl group of methanol below 35°C [21]. An FTS temperature-control system was used in variable temperature experiments. $T_2^*$ was determined from linewidth ($\Delta v$) measurements ($T_2^* = 1/(\pi \Delta v)$). A CPMG pulse sequence was used to measure $T_2$ [22]. Since the self-diffusion constants of mineral oil were not reported by other literature, they were measured by applying the common pulsed gradient NMR spin-echo techniques [23]. The strength of the gradient was first calibrated with the known self-diffusion constant of methanol at 25°C [24]. Linewidth, $T_2$ values, and self-diffusion constants were determined after data processing with Mnova. Sample preparation of $T_2$ measurements were done by filling SPIONs solution in the NMR tube. PFG experiments on mineral oil were initially problematic, presumably due to convective heating of the oil by the PFG pulse [25, 26]. To suppress the effects of convection, we put quartz wool (Acros Organics, 4 $\mu$m) in the NMR tube, and then filled the tube with pure mineral oil. We then waited until the quartz wool was thoroughly soaked before the PFG experiment. This procedure allowed us to obtain repeatable PFG results in mineral oil.

### D. ESR Sample and Measurements

8 nm, 4.5 nm, and 4 nm SPIONs were embedded in compatible solid matrices for the study of ESR thermometry. 8 and 4.5 nm hydrophilic SPIONs were immobilized in a solid polymer (TEGDMA) matrix based on a reported method with slight modification [27]. All chemicals were used as received. The procedure was as follows: synthesized SPIONs were dispersed and sonicated in 0.3 ml tetraethylene glycol dimethacrylate (TEGDMA) (Sigma, 90%) to obtain a homogeneous solution. 2,2-Azobis(2-methylpropionitrile) (AIBN) (Sigma, 98%) was then added to the solution followed by sonication. The solution was transferred to a 3 mm NMR micro tube and heated to allow polymerization (70°C, 4 hours). After 4 hours, the micro tube was cut to take out the SPIONs embedded solid polymer matrix, which was then put in the standard 4 mm quartz tube for ESR investigation.

4 nm hydrophobic SPIONs were embedded in paraffin wax (Sigma, MP: 52-58°C). Synthesized SPIONs were dispersed in melted paraffin at 60°C while stirring at 200 rpm. After the homogeneous solution was formed, the mixture solidified by removing the heat source, resulting in homogeneous paraffin wax with SPIONs homogenously embedded inside. The product was then cut to fit into the standard 4 mm quartz tube for the ESR study.

ESR measurements were conducted on a Bruker EMXPlus spectrometer operating at 9.4 GHz (X band) with a modulation frequency of 100 kHz and a modulation amplitude of 4 G. An Oxford helium gas flow cryostat was used in variable temperature experiments from 100 K to 290 K.

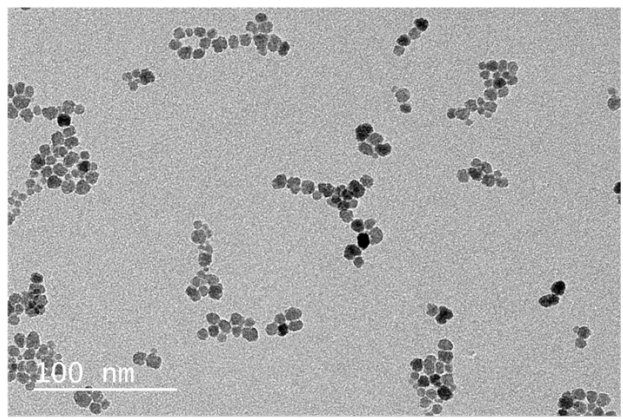

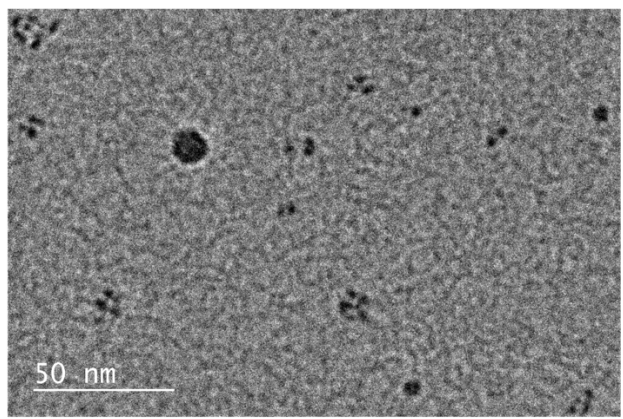

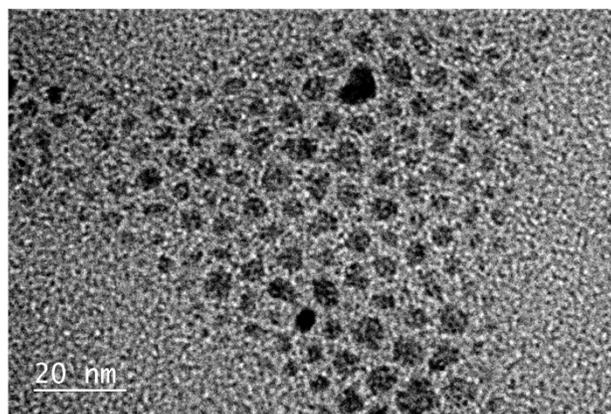

Figure 1. TEM images of (a) 8 nm (b) 4.5 nm and (c) 4 nm SPIONs.

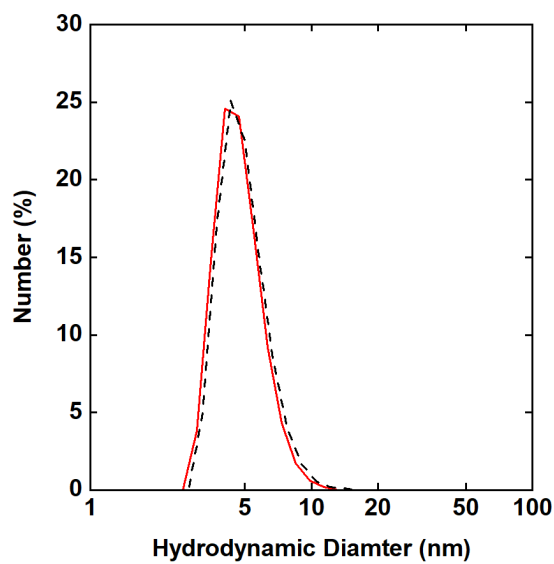

Figure 2. DLS results of 4 nm hydrophobic SPIONs used for NMR study. The hydrodynamic diameter is measured to be 4.2 nm in both hexane (black dashed line) and mineral oil (red solid line).

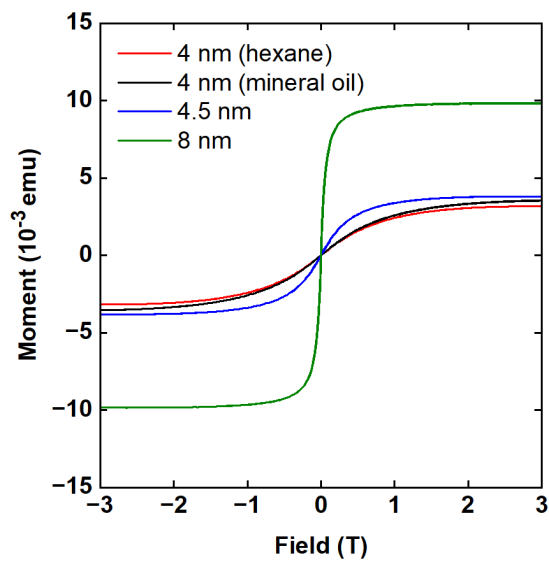

Figure 3. SQUID results of 8 nm, 4.5 nm, and 4 nm SPIONs. Measurements for 8 nm (green line) and 4.5 nm (blue line) SPIONs are performed in water at 300 K and that for 4 nm SPIONs are carried out in hexane (red line, 298 K) and mineral oil (black line, 293 K), respectively.

## III. NUCLEAR MAGNETIC RESONANCE THERMOMETRY

### A. Background

The advantages of utilizing spin-spin relaxation time ($T_2$) as a thermometer in nuclear magnetic resonance (NMR) imaging over traditional thermometers such as spin-lattice relaxation time ($T_1$) and chemical shift have been demonstrated in our previous paper [3]. $T_2$ mapping provides high-resolution images within a short acquisition time and without suffering from motional artifacts [3], making $T_2$ as a powerful thermometer in NMR thermometry. In the previous paper, the capability of $T_2$ as a thermometer in water has been demonstrated using commercially available 15 nm SPIONs [3]. Here, 4 nm SPIONs were dispersed in hexane to show a 1.4 times improvement in $T_2$ sensitivity with temperature dependent magnetization of 4 nm SPIONs in the temperature range of our study. In addition, the same SPIONs were dispersed in mineral oil to evaluate the potential of NMR $T_2$ thermometry technique in engineered fluids. We also dispersed 8 nm SPIONs in water for NMR study, which shows that $T_2$ scales with the self-diffusion constant of the water, and outer sphere theory is effective to estimate the volume fraction of SPIONs in the fluids. As there is no improvement in $T_2$ sensitivity, the results of 8 nm SPIONs in water are displayed in APPENDIX B.

To use $T_2$ as a thermometer in NMR thermometry, motional narrowing needs to be satisfied in the fluid of interest and the theory was elaborated in the previous paper [3]. Here, the same theory is applied. If motional narrowing is satisfied, outer sphere theory can be applied [3]:

$$\frac{1}{T_2} = \frac{1}{T_2^*} = \frac{5}{9} v_f \tau_c (\Delta\omega)^2 = \frac{5}{9} v_f \left(\frac{d^2}{4D}\right) \left(\sqrt{\frac{4}{5}} \frac{\gamma \mu_0 M_p}{3}\right)^2 \quad (2)$$

In the equation, $v_f$ is the volume fraction of SPIONs in the fluid, $\tau_c$ is the correlation time of protons to diffuse past the SPION, $\Delta\omega$ is the average frequency shift of a proton adjacent to the surface of the SPION. $d$ is the diameter of the SPION, $D$ is the self-diffusion constant of the fluid, $\gamma = 2.672 \times 10^8$ rad/Ts is the gyromagnetic ratio of the proton, $\mu_0 = 4\pi \times 10^{-7}$ Tm/A is the permeability of free space, $M_p$ is the magnetization of the SPION. Based on outer sphere theory, the temperature dependence of $T_2$ comes from temperature dependent magnetization of the SPIONs and temperature dependent self-diffusion of the fluid. Only the latter was utilized in our previous work [3], while here we utilize both with 4 nm SPIONs.

### B. Results and Discussion

The temperature dependent magnetization of 4 nm SPIONs in hexane and mineral oil is measured with a field sweep from -3 T to 3 T by SQUID at the same temperature points for NMR study. For hexane sample, the studied temperature points are 248 K, 273 K, 298 K, and 333 K; for mineral oil sample, they are 273 K, 293 K, 313 K, 333 K, and 353 K. Data at each temperature is then fit with modified Langevin function, see Eq. 1, to determine the magnetic moment ($\mu_p$) of the SPION at each temperature.

We expect that the total magnetic moment of a sample ($N\mu_p$) should decrease as the temperature increases with the same dependence on temperature as the magnetic moment of a particle ($\mu_p$) since the number of particles ($N$) in the sample should be constant. In Figure 4 (a), this trend can be observed for the hexane sample. Therefore, we start the fitting at 248 K with three free parameters ($N, \mu_p, \chi$) for hexane sample, see Eq. 1. After determining the best-fit values of the parameters, $N$ and $\chi$ are fixed for all other temperature points to obtain the best-fit values of $\mu_p$. For the mineral oil sample, Figure 4 (b) reveals that $N\mu_p$ does not vary with temperature in the same way as $\mu_p$ at 313 K, 333 K, and 353 K. This indicates that $N$ seen by the instrument is not identical at different temperature points, and hence we allow $N$ to vary to find the best-fit values of $\mu_p$ at 313 K, 333 K, and 353 K. By doing this, the fit values of $N$ increase slightly by a factor of 1.1 in the temperature range of measurement (273 K - 353 K). On the other hand, if we still use $N$ as a fixed parameter at these three temperature points, we find significant discrepancy in the temperature dependence of $\mu_p$ between hexane and mineral oil sample (see Appendix D). Figure 5 summarizes the best-fit values of magnetic moment at each temperature.

In addition to the possible reduction of Curie temperature [7], the enhanced temperature dependence of magnetic moment (or magnetization) of 4 nm SPIONs might also be associated with the defect structures inside the particles. The magnetic moment of 4 nm SPIONs is suppressed by a factor of two compared with 4.5 nm SPIONs. The 4.5 nm SPIONs were synthesized from Fe(acac)$_3$ at 230°C for 2 hours, which is more likely to result in low defect densities without spin-canted structure [28]. On the other hand, 4 nm SPIONs were synthesized from iron oleate at 200°C for 0.5 hours, which tends to introduce defects such as antiphase boundaries (APBs) and a spin-canted layer [28]. APBs are often introduced in the particle interior due to topotaxial oxidation from FeO phase to Fe$_3$O$_4$ or Fe$_2$O$_3$ phase during thermal decomposition from iron oleate [28-30]. The multi-phase and APB structures could lead to reduced magnetization [28-31]. Therefore, we suggest that the origin of the stronger temperature dependent magnetization of our 4 nm SPIONs is related to the internal defect structures of the particle in addition to the possible reduction of the Curie temperature created by the large surface to volume ratio.

The temperature dependence of magnetic moment, see Figure 5, is equivalent to the temperature dependence of the magnetization since the volume of the SPION is not significantly temperature dependent. Here, we use $DM^{-2}$ to represent the temperature dependence of diffusion constant ($D$) combined with that of magnetization ($M$) in the following discussion for NMR results. To better describe the temperature sensitivity near room temperatures, we introduce a sensitivity coefficient, $\xi_S^T$, which provides the percentage change in signal $S$ for 1% change in the absolute temperature $T$ [4]. To perform the calculation of sensitivity, we fit the data for temperature dependence of $D$, $DM^{-2}$, and $T_2$ to an Arrhenius type of fit, and the details of the fit can be found in APPENDIX C.

The temperature dependence of spin-spin relaxation time ($T_2$) in hexane in the presence of 4 nm SPIONs between 248 K and 333 K is shown in Figure 6 (a). Based on the self-diffusion constant of CH$_4$ group proton of hexane [32], the temperature sensitivity ($\xi_D^T$) is 3.48 at room temperature. Combined with temperature dependent magnetization of 4 nm SPIONs, the sensitivity of $DM^{-2}$ is enhanced to 4.54 and that of $T_2$ is 4.96, which is a 1.4 times improvement from the temperature dependence of self-diffusion alone. The close values between $T_2$ and $DM^{-2}$ sensitivity indicate that motional narrowing is in effect ($\Delta\omega\tau_c \leq 1$), which can also be confirmed in Figure 6 (a) where $T_2$ scales with $DM^{-2}$. For 15 nm SPIONs in water [3], it was shown that $T_2^*$ is equivalent to $T_2$ when the linewidth induced by SPIONs is much larger than the natural linewidth of the liquid proton. Here, $T_2^*$ deviates from $T_2$ for all volume fractions because the relaxation induced by inhomogeneities in the static field due to imperfect shimming is comparable to the relaxation induced by SPIONs, which itself is quite small compared with the relaxation induced by 15 nm SPIONs in water [3]. However, $T_2$, not affected by static field inhomogeneities, remains an effective thermometer without suffering from shimming problems in our measurement.

The critical diameter ($d_c$) of the SPIONs that satisfies motional narrowing ($\Delta\omega\tau_c \leq 1$) is defined in our previous paper [3], which gives:

$$d_c = \left(\sqrt{\frac{4}{5}} \frac{\gamma \mu_0 M_p}{12D}\right)^{-\frac{1}{2}} \quad (3)$$

Here, $D = 4.21 \times 10^{-9}$ m²/s is the self-diffusion constant of hexane at 25°C [32], and $M_p = 3.3 \times 10^5$ A/m of 4 nm SPIONs from SQUID are used for calculation of $d_c$, which is determined to be 22 nm. In Figure 2, the DLS result shows that the hydrodynamic diameter of 4 nm SPIONs in hexane is 4.2 nm. Therefore, the temperature dependence of $T_2$ in hexane scales expectedly with the self-diffusion of hexane combined with temperature dependent magnetization of 4 nm SPIONs.

With the measured $T_2$ value at room temperature, we can estimate the volume fraction of SPIONs in hexane from outer sphere theory based on Eq. 2. Here, we compare the estimated volume fraction from outer sphere theory and SQUID measurement. The volume fraction ($v_f$) of 4 nm SPIONs in hexane used for SQUID measurement is 120 ppm ($v_f = v_{p,total}/v_{fluid} = N\mu_p/(M_p v_{fluid})$). After diluted by two times, the ferrofluid is then used for $T_2$ measurement. Therefore, in $T_2$ measurement, the volume fraction of the most concentrated sample from outer sphere theory should be close to 60 ppm if motional narrowing is fully satisfied. From outer sphere theory and $T_2$ measurement, the volume fraction is calculated to be 52 ppm, which is close to 60 ppm. This indicates that a majority of the SPIONs participate in motional narrowing since the hydrodynamic size of them is smaller than the critical diameter. In APPENDIX B, we show that outer sphere theory is also effective in estimating the volume fraction of 8 nm SPIONs in water. In Figure 7, it is shown that $T_2$ scales inversely linear with the volume fraction of SPIONs in hexane, as expected from outer sphere theory [3]. We expect the SPIONs remain stable during NMR measurements as the magnetic interaction between SPIONs is negligible even for the maximum volume fraction of 52 ppm [3].

After these results in hexane, we proceeded to study $T_2$ sensitivity in mineral oil, which is commonly used as a heat transfer fluid [33]. From DLS and SQUID magnetometry, we expect the 4 nm SPIONs in mineral oil are identical as that in hexane. We study temperature range between 273 K and 353 K. The results of 4 nm SPIONs in mineral oil are shown in Figure 6 (b). From Arrhenius fit of the temperature dependence, the sensitivity of $D$ of mineral oil is 10.38. Combined with temperature dependent magnetization of 4 nm SPIONs, the sensitivity of $DM^{-2}$ is improved to be 11.86 and that of $T_2$ is 11.62. From Eq. 3, the critical diameter ($d_c$) satisfying motional narrowing in mineral oil was calculated to be 2.2 nm at 25°C; Here, $D = 4.21 \times 10^{-11}$ m²/s is the self-diffusion constant of mineral oil at 25°C and $M_p = 3.4 \times 10^5$ A/m of 4 nm SPIONs from SQUID are used for calculation of $d_c$. In Figure 2, DLS result shows that the hydrodynamic diameter of 4 nm SPIONs in mineral oil is 4.2 nm. Hence, a reduction in the temperature dependence of $T_2$ in mineral oil might be expected as it was reported that the temperature dependence could decrease due to particle aggregation [34]. However, in Figure 6 (b), it shows that $T_2$ still scales with $DM^{-2}$. As the hydrodynamic size is not significantly larger than the critical size, motional narrowing is partially in effect, and therefore $T_2$ still exhibits remarkable temperature dependence. The partially satisfied motional narrowing can be confirmed by the volume fraction of SPIONs estimated by the outer sphere as well.

The volume fraction of 4 nm SPIONs in mineral oil used for SQUID measurement is 120 ppm ($v_f = v_{p,total}/v_{fluid} = N\mu_p/(M_p v_{fluid})$). After diluted by two times, the ferrofluid is then used for $T_2$ measurement. Therefore, in $T_2$ measurement, the volume fraction of the most concentrated sample from outer sphere theory should be close to 60 ppm if motional narrowing is fully satisfied. However, the largest volume fraction is calculated to be only 8 ppm based on outer sphere theory, which indicates that there is only a small fraction of SPIONs with hydrodynamic diameter smaller than critical diameter and participating in the motional narrowing. Despite this small fraction (8 ppm) of SPIONs involved in motional narrowing, the temperature dependence of $T_2$ doesn't decrease and remains the same for lower concentration samples. In Figure 7, it is also shown that $T_2$ scales inversely linear with the volume fraction of 4 nm SPIONs in mineral oil, consistent with outer sphere theory. Based on the result of mineral oil, we find, as long as the hydrodynamic diameter of the particle is not significantly larger than the critical diameter and there is a fraction of particles involved in motional narrowing, the temperature dependence of $T_2$ can still scale with $DM^{-2}$.

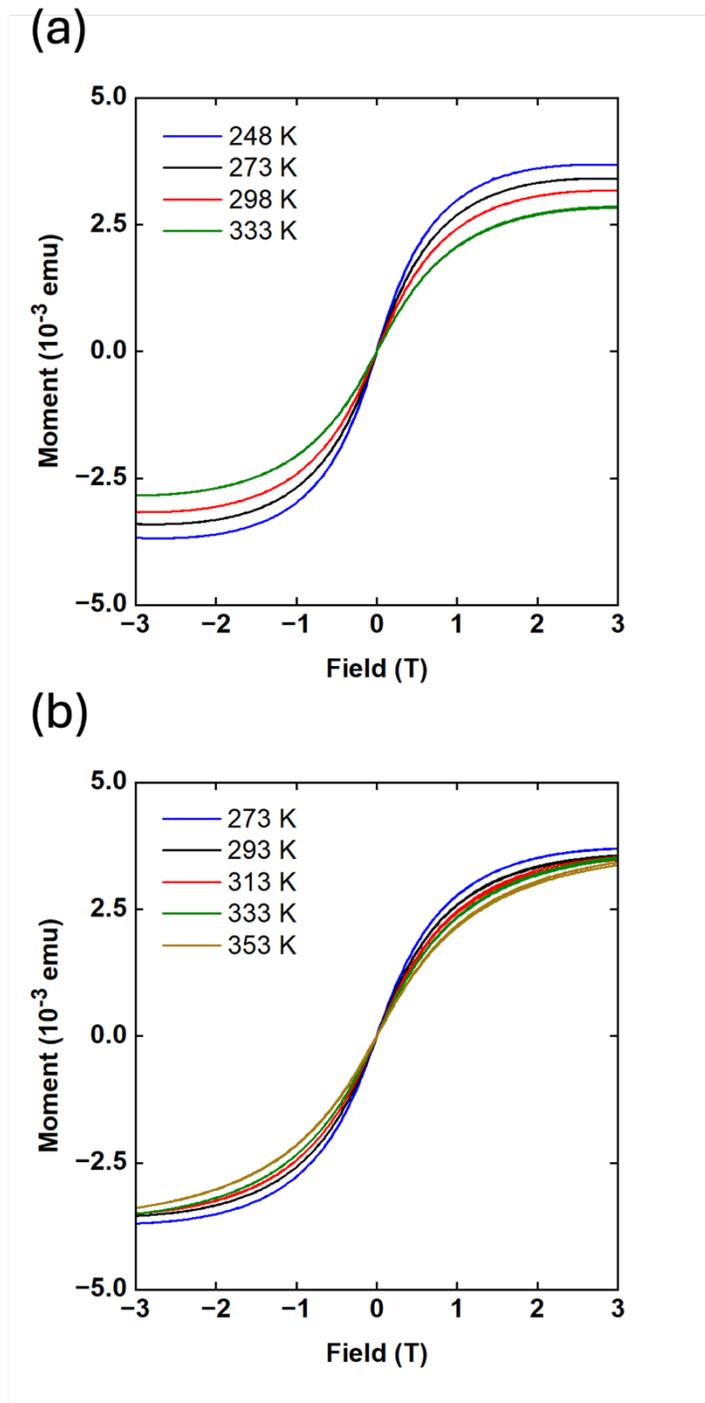

Figure 4. Field sweep results of 4 nm SPIONs in (a) hexane and (b) mineral oil at different temperature points. Data is then fit with the modified Langevin function, see Eq. 1, to find the magnetic moment ($\mu_p$) of the SPION at each temperature point, which is plotted in Figure 5 to show the temperature dependence.

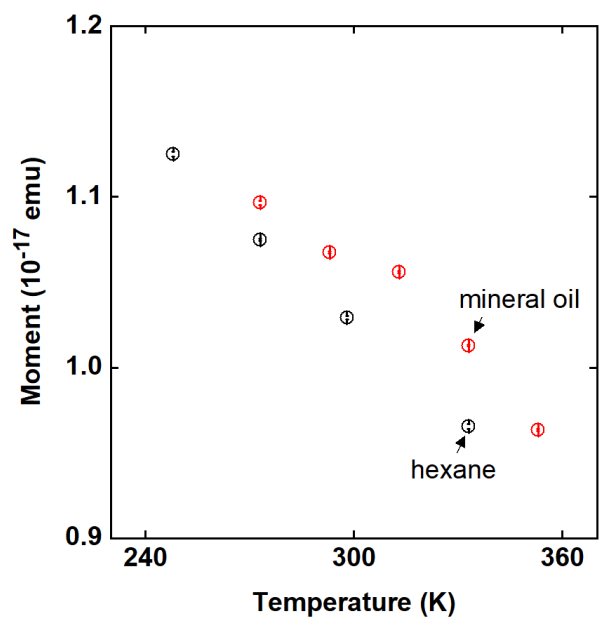

Figure 5. The temperature dependence of magnetic moment ($\mu_p$) in hexane (black) and mineral oil (red). The error bars represent 95% confidence interval of fitting, which is comparable to the size of the symbols in the figure.

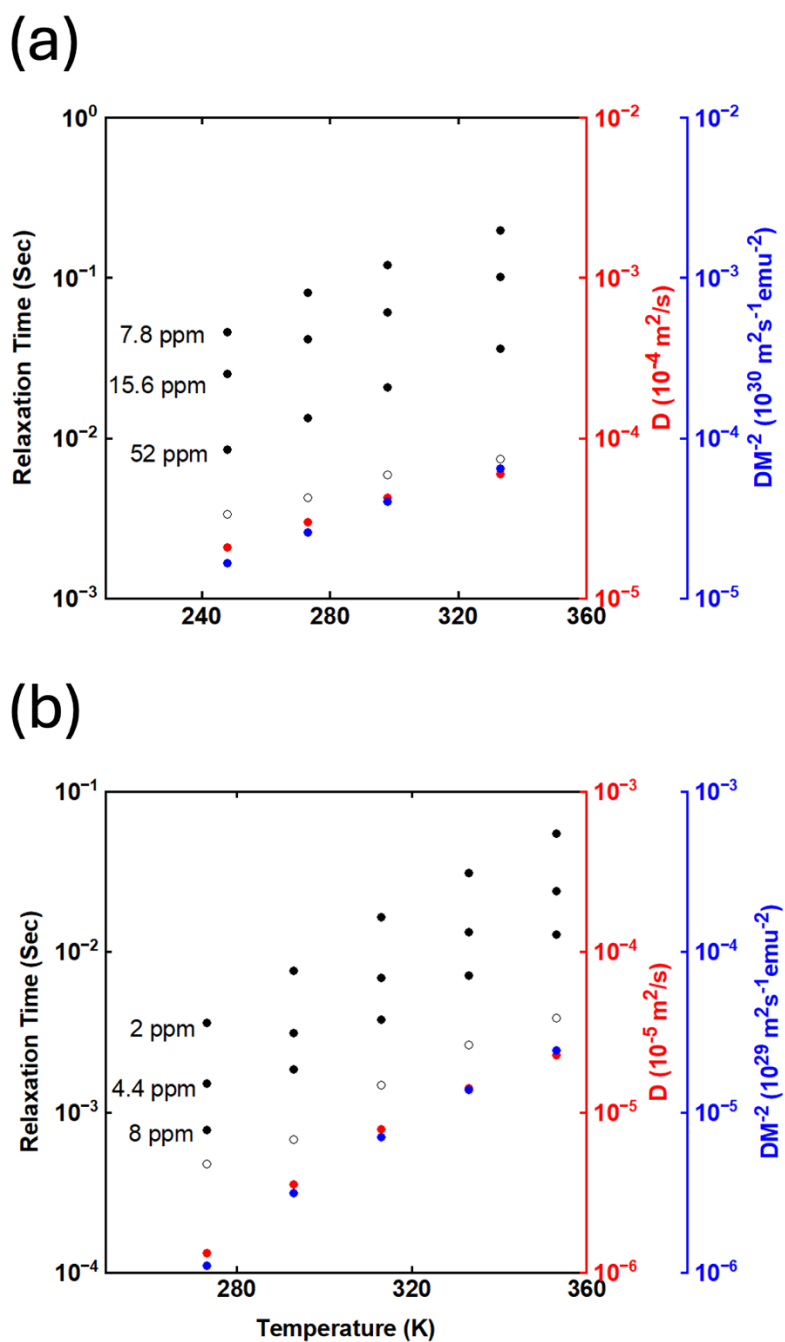

Figure 6. Temperature dependence of $T_2$ (black filled circles) and $T_2^*$ (black open circles) with 4 nm SPIONs in (a) hexane and (b) mineral oil. Volume fractions in the unit of ppm are estimated from outer sphere theory. The red circles are self-diffusion constant ($D$) of the fluid and the blue circles are self-diffusion constant combined with the magnetization of 4 nm SPIONs ($DM^{-2}$). $D$ of hexane in (a) were measured by Harris [32].

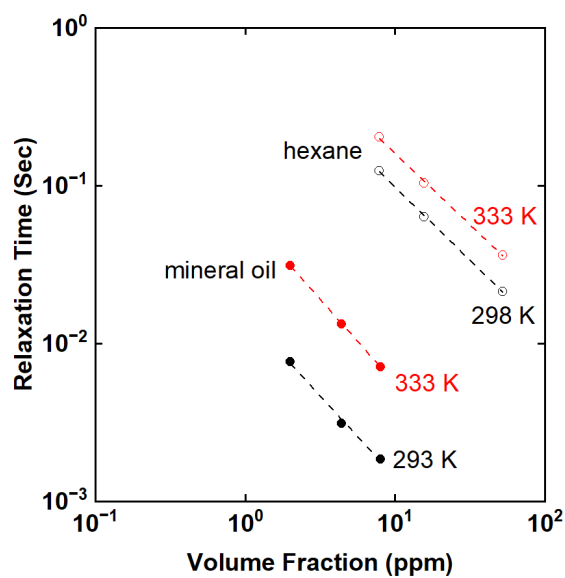

Figure 7. Dependence of relaxation time ($T_2$) on volume fraction in hexane (open circles) and mineral oil (filled circles) at room temperature (black) and 333 K (red). The relaxation time ($T_2$) scales inversely linear with the volume fraction of 4 nm SPIONs in both hexane and mineral oil, consistent with outer sphere theory.

**IV. ELECTRON SPIN RESONANCE THERMOMETRY**

**A. Background**

ESR linewidth of the SPIONs was reported to exhibit strong temperature dependence [10-13]. Here, the potential of utilizing ESR linewidth to probe temperature fields in solids is investigated, where NMR thermometry cannot be applied. The focus is to study SPION size and concentration effects on the temperature dependence of linewidth.

**B. Results and Discussion**

SPIONs in a non-magnetic solid matrix can be regarded as a system consisting of randomly oriented particles and distributed anisotropy fields [10]. At low temperatures, this spread of anisotropy fields leads to a wide range of resonant frequencies, resulting in inhomogeneous broadening of linewidth. As the temperature increases, the effective anisotropy is reduced by thermal fluctuation and causes the linewidth to decrease [10-13]. Therefore, at higher temperatures, the line shape becomes narrower and more symmetrical.

The temperature dependence of linewidth ($\Delta H$) can be fit with [11-13]:

$$\Delta H = H_T L\left(\frac{\mu_p H}{k_B T}\right) = H_0 G(y) L\left(\frac{\mu_p H}{k_B T}\right) \quad (4)$$

Here, $H_T$ is the saturation line at a temperature $T$, $L\left(\frac{\mu_p H}{k_B T}\right) = \coth\left(\frac{\mu_p H}{k_B T}\right) - \frac{k_B T}{\mu_p H}$ is the Langevin function, $\mu_p$ is the magnetic moment of a particle, and $H = 3400$ G is the resonance field. The temperature dependence of $H_T$ arises from thermal fluctuation-induced variation of anisotropy and can be written as $H_T = H_0 G(y)$, where $H_0$ is the saturation linewidth at 0 K and $G(y)$ is the superparamagnetic averaging factor given as

$$G(y) = \frac{1}{L(y)} - \frac{10}{y} + \frac{35}{y^2 L(y)} - \frac{105}{y^3} \quad (5)$$

Here, $y = KV_s/k_B T$; $K$ is the anisotropy constant and $V_s$ is the reference volume (presumably the greatest volume in the statistical ensemble). $L(y) = L(KV_s/k_B T)$ is the Langevin function. We follow a common practice to fit the data where $H_0$ and $V_s$ are the free parameters while $\mu_p$ and $K$ are the fixed parameters [11-13]. The values of $\mu_p$ are determined from the fitting of Langevin function (Eq. 1) to the data of SQUID (Figure 3). The values of $K$ are estimated from the expression [35,36]:

$$K = K_v + (6\emptyset/D)K_s \quad (6)$$

Where $K_v = 13$ kJ/m³ [11, 36] is the bulk anisotropy constant of magnetite and $K_s = 2.8 \times 10^{-5}$ J/m² [35] is the surface anisotropy constant; $6\emptyset/D$ ($D$: diameter) is the surface to volume ratio. Since our SPIONs are nearly spherical, $\emptyset$ is equal to 1 [35, 36]. Therefore, for 4 nm and 4.5 nm particles, $K$ is estimated to be 55 kJ/m³ and 50 kJ/m³, respectively. For 8 nm SPIONs, bulk value (13 kJ/m³) is used since it was found that $K$ reaches asymptotically the bulk value from $D > 8$ nm [36].

The best-fit values of all the parameters are summarized in Table 1. Based on the fit values, we find that there are two critical parameters that determines the temperature dependence of linewidth: the reference value ($V_s$) and the magnetic moment ($\mu_p$) of the particle. The temperature dependence of linewidth for 4 nm SPIONs scales with $T^{-2}$ while that of 4.5 nm SPIONs scales with $T^{-1}$. Despite the similar size between them, the magnetic moment of 4.5 nm SPIONs is a factor of two larger than that of 4 nm SPIONs. Moreover, the fit values of $V_s$, i.e., the greatest volume in the statistical ensemble, reveal that there is a broader size distribution for 4.5 nm SPIONs ($V_s = 892$ nm³) than 4 nm SPIONs ($V_s = 181$ nm³). The difference in the size distribution lies in the nature of different synthetic method of SPIONs. For 8 nm SPIONs, the temperature dependence of linewidth is not significant as $V_s$ and $\mu_p$ are much larger than that of smaller particles. Therefore, to enhance temperature dependence of ESR linewidth, we find that it is important to reduce the particle magnetic moment and the overall volume of the particles.

In addition to the size effects on the temperature dependence of ESR signal, we also study the concentration effect of the SPIONs in the matrices. For each size of SPIONs, two concentrations of the same-sized particles in the same solid matrix were prepared. We took care to fix the dimension of solid samples of different concentrations, so the region of the samples seen by the instrument is consistent. The lower concentrated sample is approximately half of the higher one.

This can be confirmed by the double integrated intensity of the signal, which is proportional to the number of spins in the sample at the same temperature. In Figure 8 and 9, our results show that a change in concentration does not change the signal behavior, meaning that the effects of dipole interaction between the particles can be neglected, which was also observed in another study [37].

We also notice that there is weak bump in the signal at $H = 1500$ G for all samples, which results from paramagnetic $Fe^{3+}$ ion residue [38]. The intensity of residue signal grows as cooling and eventually merges with the inhomogeneous broadened linewidth of SPIONs. Based on the location of the $Fe^{3+}$ signal, it can be assured that it doesn't affect the superparamagnetic signals coming from SPIONs at $H = 3400$ G.

| Sample | Number of Spins | Size (nm) | $\mu_p$ (J/T) | $H_0$ (T) | $V_s$ (nm$^3$) | $K_1$ (kJ/m$^3$) |
|---|---|---|---|---|---|---|
| A | $1.8 \times 10^7$ | 8.1 | $1.37 \times 10^{-19}$ | 0.188 | 7867 | 13 |
| B | $10^7$ | | | | | |
| C | $5.1 \times 10^6$ | 4.5 | $2.26 \times 10^{-20}$ | 0.179 | 892 | 50 |
| D | $2.7 \times 10^6$ | | | | | |
| E | $7.7 \times 10^6$ | 3.9 | $1.07 \times 10^{-20}$ | 0.603 | 181 | 55 |
| F | $4.1 \times 10^6$ | | | | | |

Table 1. Summaries of ESR solid sample properties. Number of spins in the sample is obtained from double integrated intensity of the signal. The concentration of spins in sample (A),(C),(E) is approximately a factor of two higher than that in sample (B),(D),(F). The magnetic moment ($\mu_p$) of 8 nm and 4 nm SPIONs here is identical to that in the fluids used for NMR study. The values of $H_0$ and $V_s$ are the best-fit values by Eq. 4 and $K_1$ are estimated by Eq. 6.

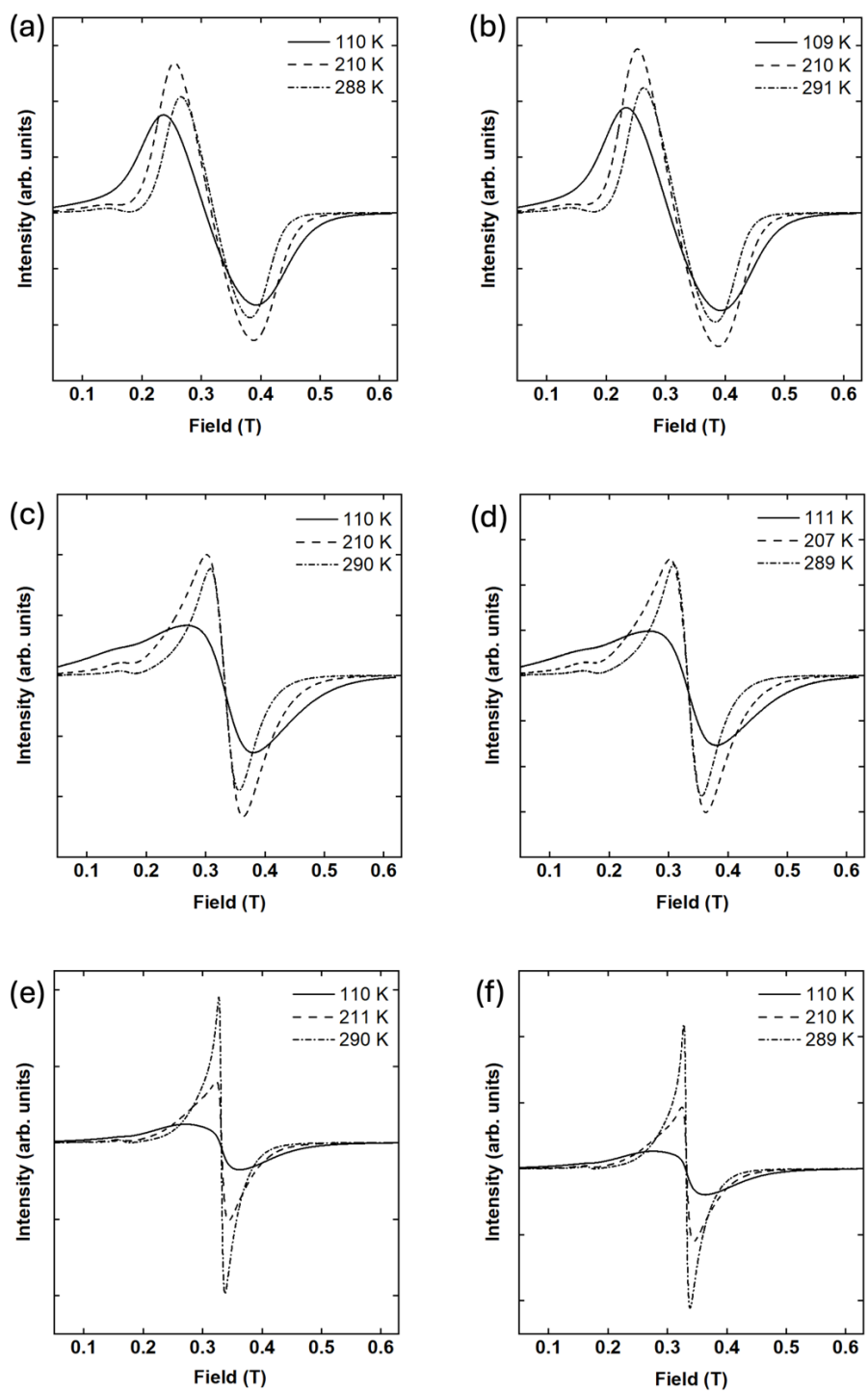

Figure 8. ESR spectra of (a),(b) 8 nm SPIONs in polymer, (c),(d) 4.5 nm SPIONs in polymer, and (e),(f) 4 nm SPIONs in paraffin wax at 110 K, 210 K, and 290 K. Linewidth is narrower at higher temperatures and decreases with the size of the SPIONs. The concentration of SPIONs in sample (a),(c),(e) is approximately twofold higher than sample (b),(d),(f). It is shown that concentration does not make noticeable difference in the spectra.

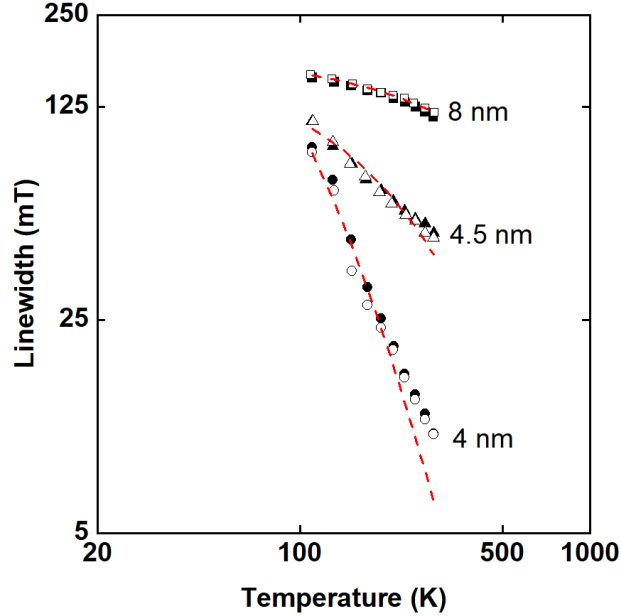

Figure 9. Temperature dependence of ESR linewidth. The filled legends are approximately twofold more concentrated than the open legends. The concentration of SPIONs does not affect the linewidth. The red dash line are the fit data from Eq. 4.

## V. CONCLUSION

NMR results confirm that $T_2$ is an extremely effective thermometer in the presence of SPIONs without suffering from the problem of shimming at low concentrations in fluids. For hexane, $T_2$ sensitivity ($\xi_{T_2}^T = 4.96$) is improved with temperature dependent magnetization of 4 nm SPIONs by 1.4 times from self-diffusion ($\xi_D^T = 3.48$) alone between 248 K and 333 K. For mineral oil, though motional narrowing is not fully satisfied with 4 nm SPIONs due to the slow diffusion of mineral oil, $T_2$ also exhibits high sensitivity ($\xi_{T_2}^T = 11.62$) as the temperature dependence of self-diffusion in mineral oil is already quite significant ($\xi_D^T = 10.38$) between 273 K and 353 K. Based on the results in hexane and mineral oil, the applications of NMR $T_2$ thermometry in engineered fluids are promising, which may help to solve heat management problems in electronic devices and turbine engine systems.

ESR results show that linewidth exhibits a significant temperature dependence, with the strongest following a $T^{-2}$ for 4 nm SPIONs in paraffin wax. To enhance the temperature dependence of ESR linewidth, it is important to use small particle volume with reduced magnetic moment. While ESR linewidth is a sensitive thermometer, the linewidth we show may be too broad to be applicable in 3D imaging. Due to the limitations of our instrument, we couldn't do measurements above room temperature. However, as the linewidth decreases with temperatures, it may be expected to be a useful thermometer above room temperatures where the linewidth becomes narrow enough for 3D imaging with small enough magnetic nanoparticles. For low temperature applications, utilizing different types of superparamagnetic nanoparticles that have narrower linewidth may be a good strategy for 3D imaging.

Combining our results of NMR and ESR, the size effect is found to be a critical role in improving the sensitivity of both thermometry techniques. In general, the smaller the SPION is, the higher temperature sensitivity it can provide with an appropriate thermometry. Therefore, to develop highly sensitive magnetic resonance techniques, using small SPIONs is beneficial.

**ACKNOWLEDGEMENTS**

The authors are grateful to Dr Toby Woods of the School of Chemical Sciences, University of Illinois for assistance with the ESR measurements. The authors also want to thank Dr Lingyang Zhu of the School of Chemical Sciences, University of Illinois for assistance with the NMR measurements. The Quantum Design MPMS 3 SQUID Magnetometer used in the experiment was partially funded through Illinois MRSEC NSF Award Number DMR-1720633. Major funding for Bruker EMXPlus was provided by the National Science Foundation Award 1726244 (2017) to the School of Chemical Sciences EPR Lab at the University of Illinois.

# APPENDIX A

# DYNAMIC VISCOSITY OF MINERAL OIL

To ensure accurate DLS measurements of 4 nm SPIONs in mineral oil, the dynamic viscosity of mineral oil at 25°C is measured with a TA-DHR3 rheometer. An upper cone geometry (40 mm, 1°) is used for the measurements. Temperature is stabilized with a Peltier plate. A Shear rate in the range of 1 to 1000 $s^{-1}$ is applied to measure the dynamic viscosity, which is determined to be 18.4 cp at 25°C, as shown in Figure A.1.

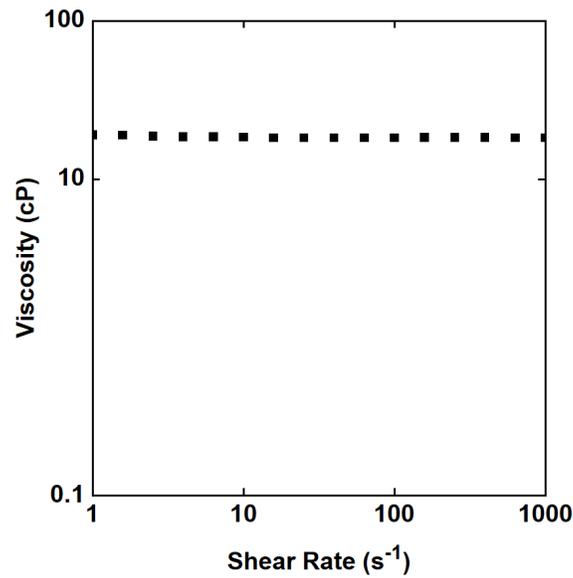

Figure A.1. The dynamic viscosity of mineral oil as a function of shear rate at 25°C.

# APPENDIX B

# VOLUME FRACTION OF SPIONS

Based on Langevin fitting, the magnetization of 8 nm hydrophilic SPIONs is determined to be $4.93 \times 10^5$ A/m. The volume fraction ($v_f$) of SPIONs in water (water volume: 100 $\mu$l) for SQUID measurement can be estimated: $v_f \times 10^{-7} = \frac{N\mu_p}{M_p} = \frac{9.86 \times 10^{-6}}{4.93 \times 10^5}$, which gives $v_f$ to be 200 ppm. The same volume fraction of 8 nm SPIONs in water was used for $T_2$ measurement. In Figure B.1, DLS result shows that the hydrodynamic diameter of 8 nm SPIONs in water is 10 nm, smaller than the critical diameter (14 nm), and therefore the motional narrowing is satisfied. In Figure B.2 (a), temperature dependence of $T_2$ of 8 nm SPIONs in water scales with self-diffusion of water [21], indicating that motional narrowing is satisfied, and outer sphere theory can be applied. The volume fraction is determined to be 215 ppm from outer sphere theory, which is close to that of SQUID. This indicates that outer sphere theory could be used to estimate the volume fraction of SPIONs in liquid samples for NMR measurements. In Figure B.2 (b), the relaxation time scales inversely linear with the volume fraction of SPIONs, as expected from the outer sphere theory.

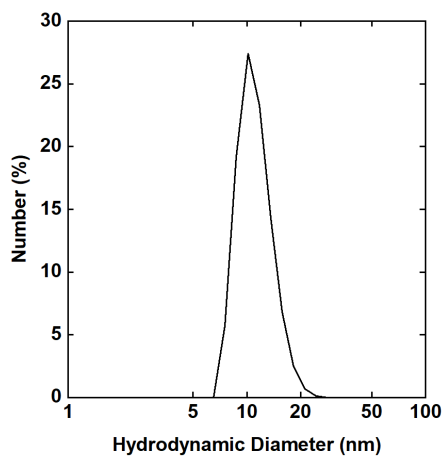

Figure B.1. DLS result of 8 nm SPIONs used for NMR study.
The hydrodynamic diameter is measured to be 10 nm in water.

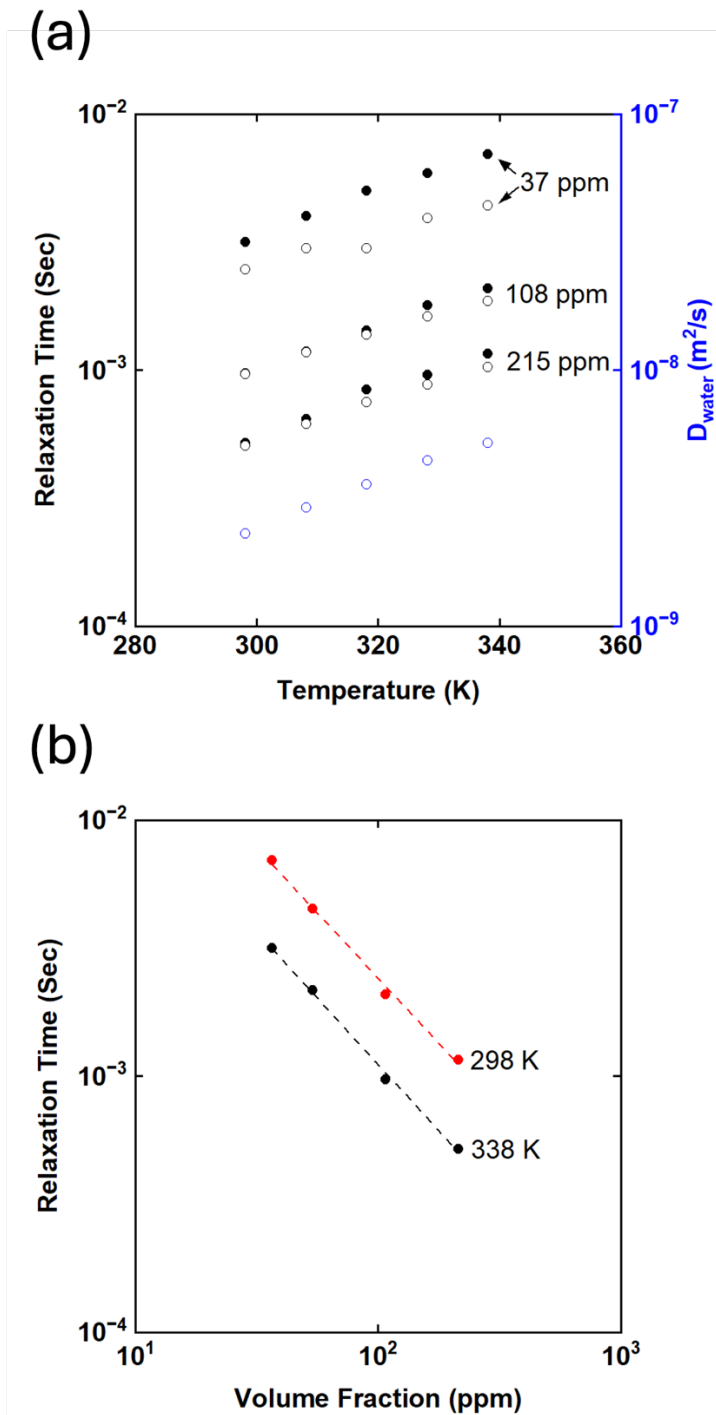

Figure B.2. (a) Temperature dependence of $T_2$ (black filled circles) and $T_2^*$ (black open circles) with 8 nm SPIONs in water. The blue open circles are the self-diffusion constant of water measured by Holz *et al.* [23]. Volume fractions are calculated from outer sphere theory. (b) $T_2$ scales inversely linear with the volume fraction of 8 nm SPIONs in water, as expected from the outer sphere theory.

# APPENDIX C

# TEMPERATURE SENSITIVITY COEFFICIENT

For NMR $T_2$ thermometry, we define a temperature sensitivity coefficient, $\xi_S^T$, to better describe and compare the sensitivity between self-diffusion constant ($D$) of the fluids and $T_2$. $\xi_S^T$ defines the percentage change in signal $S$ for 1% change in the absolute temperature $T$ [4]. For $D$, $\xi_D^T = d \ln D / d \ln T$; for $T_2$, $\xi_{T_2}^T = d \ln T_2 / d \ln T$; and for $DM^{-2}$, $\xi_{DM^{-2}}^T = d \ln DM^{-2} / d \ln T$ [39]. We expect that the self-diffusion in fluid follows Arrhenius law, which fits well with our data, as shown in Figure C.1. To obtain a good estimate of $\xi_S^T$ near room temperatures, we fit data of $D$, $T_2$, and $DM^{-2}$ as a function of $1/T$ with Arrhenius equation. Taking $D$ as an example:

$$\xi_D^T = \frac{d \ln D}{d \ln T} = \frac{T}{D} \frac{dD}{dT} \quad \text{(C.1)} \quad \text{and} \quad D = D_0 \exp\left(-\frac{E_a}{k_B T}\right) \quad \text{(C.2)}$$

Substituting $D$ into the equation of $\xi_D^T$, we obtain $\xi_D^T = E_a/k_B T$. The above equations can also be applied to $T_2$ and $DM^{-2}$. Therefore, by plotting $D$, $T_2$, and $DM^{-2}$ as a function of $1/T$ and do a mono-exponential fit, we get the constant $E_a/k_B$, which is then divided by $T = 293$ K for mineral oil and $T = 298$ K for hexane to determine sensitivity near room temperature.

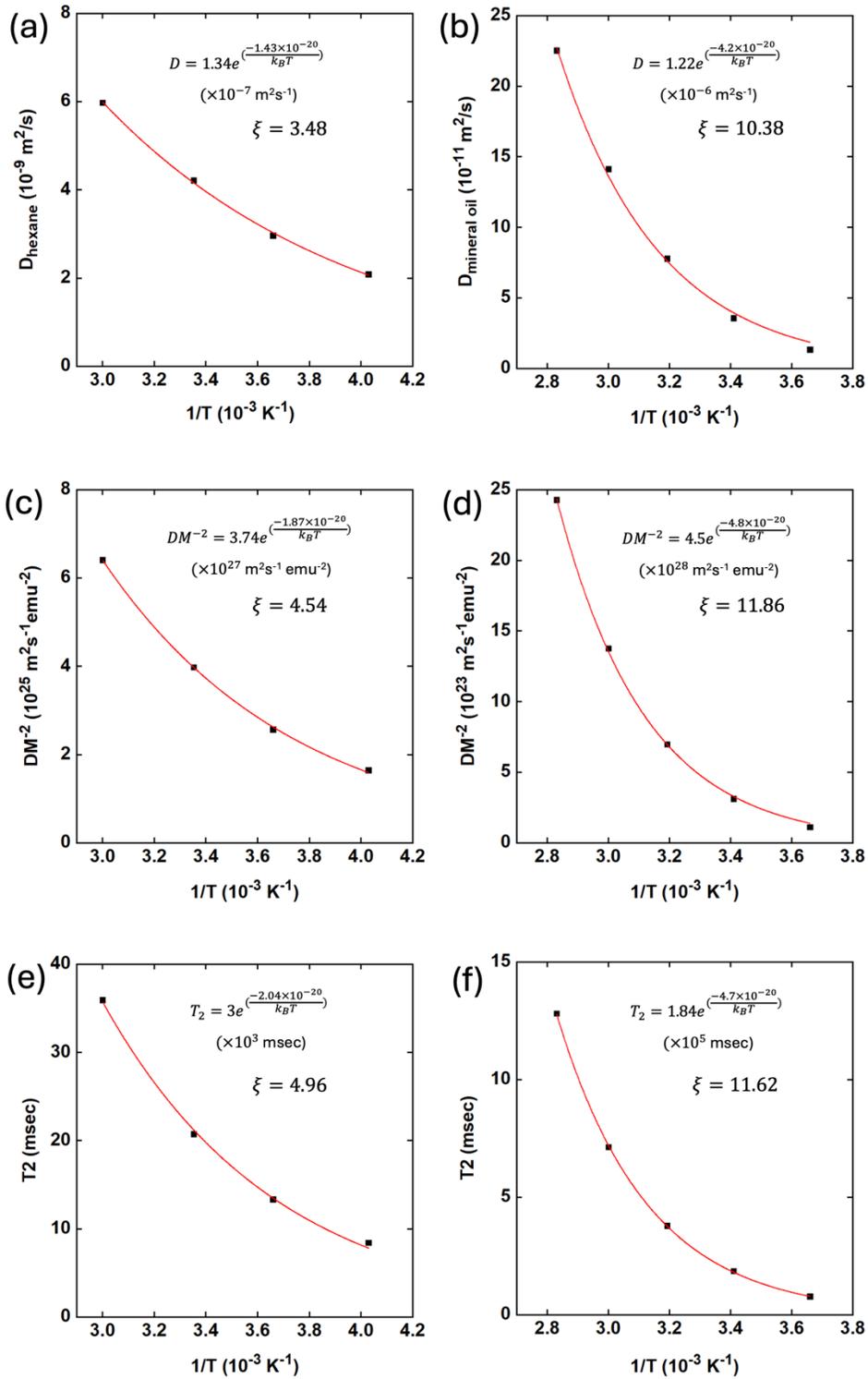

Figure C.1. Data (black square) can be fit well with Arrhenius law (red line): $A = A_0 \exp\left(-\frac{E_a}{k_B T}\right)$. Figure (a), (c), and (e) are $D$ (measured by Harris [32]), $DM^{-2}$, and $T_2$ in hexane, respectively. (b), (d), and (f) are $D$, $DM^{-2}$, and $T_2$ in mineral oil, respectively.

# APPENDIX D

# LANGEVIN FITTING OF FIELD SWEEP DATA

Figure D.1 shows the Langevin fitting of 4 nm SPIONs in hexane at 248 K, 273 K, 298 K, and 333 K. Figure D.2 shows the Langevin fitting of 4 nm SPIONs in mineral oil at 273 K, 293 K, 313 K, 333 K, and 353 K. The fitting of the data is based on:

$$\mu_{SQUID} = N\mu_p \left[\coth\left(\frac{\mu_p H}{k_B T}\right) - \left(\frac{k_B T}{\mu_p H}\right)\right] + \chi H \quad (D.1)$$

where $\mu_{SQUID}$ is the data measured by SQUID magnetometry; $N\mu_p$ is the total magnetic moment of the sample where $N$ is the number of particles in the sample and $\mu_p$ is the magnetic moment of a particle. The liner term $\chi H$ is added to take diamagnetic background into account [33]. In the fitting of hexane, $N$ and $\chi$ are fixed parameters to find the best-fit values of $\mu_p$ for all temperatures. In the fitting of mineral oil, $\chi$ is a fixed parameter to find the best-fit values of $\mu_p$ and $N$. Figure D.3 shows the results of fixing $N$ in the fitting at 313 K, 333 K, and 353 K for mineral oil, leading to discrepancy of temperature dependence of magnetic moment between mineral oil and hexane.

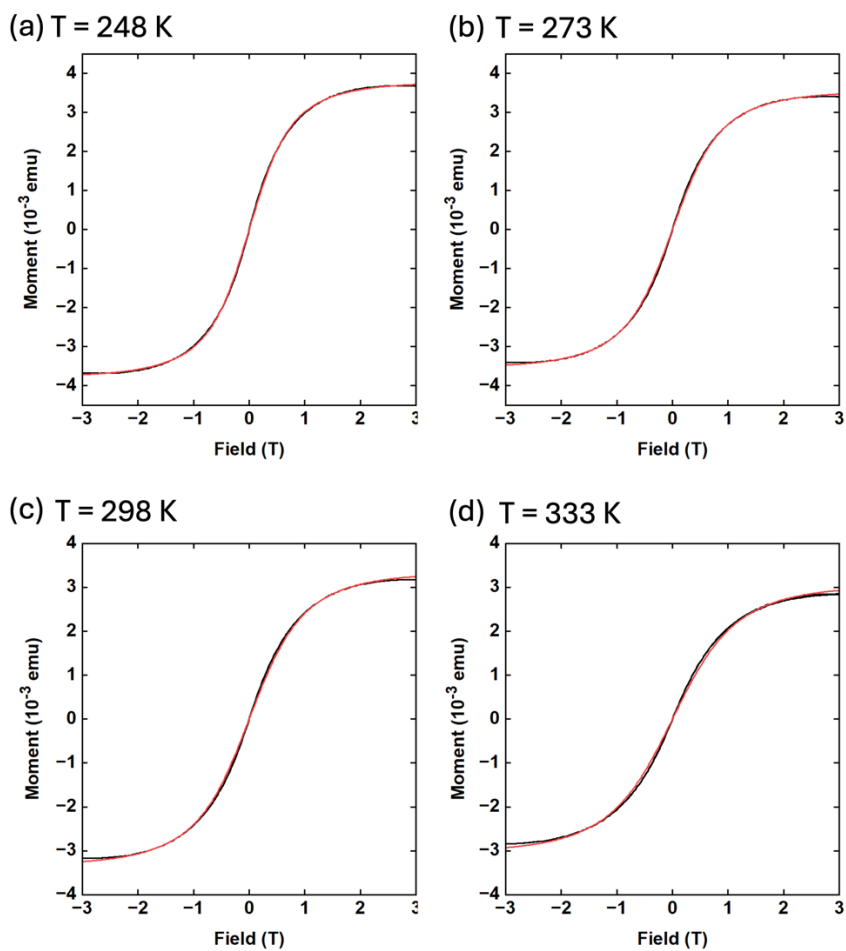

Figure D.1. Data (black line) is fit with modified Langevin function (red line). The best-fit value of $N$ is $3.95 \times 10^{14}$ and that of $\chi$ is $9.3 \times 10^{-8}$ J/T². The 95 % CI stands for 95 % confidence interval of the fitting.

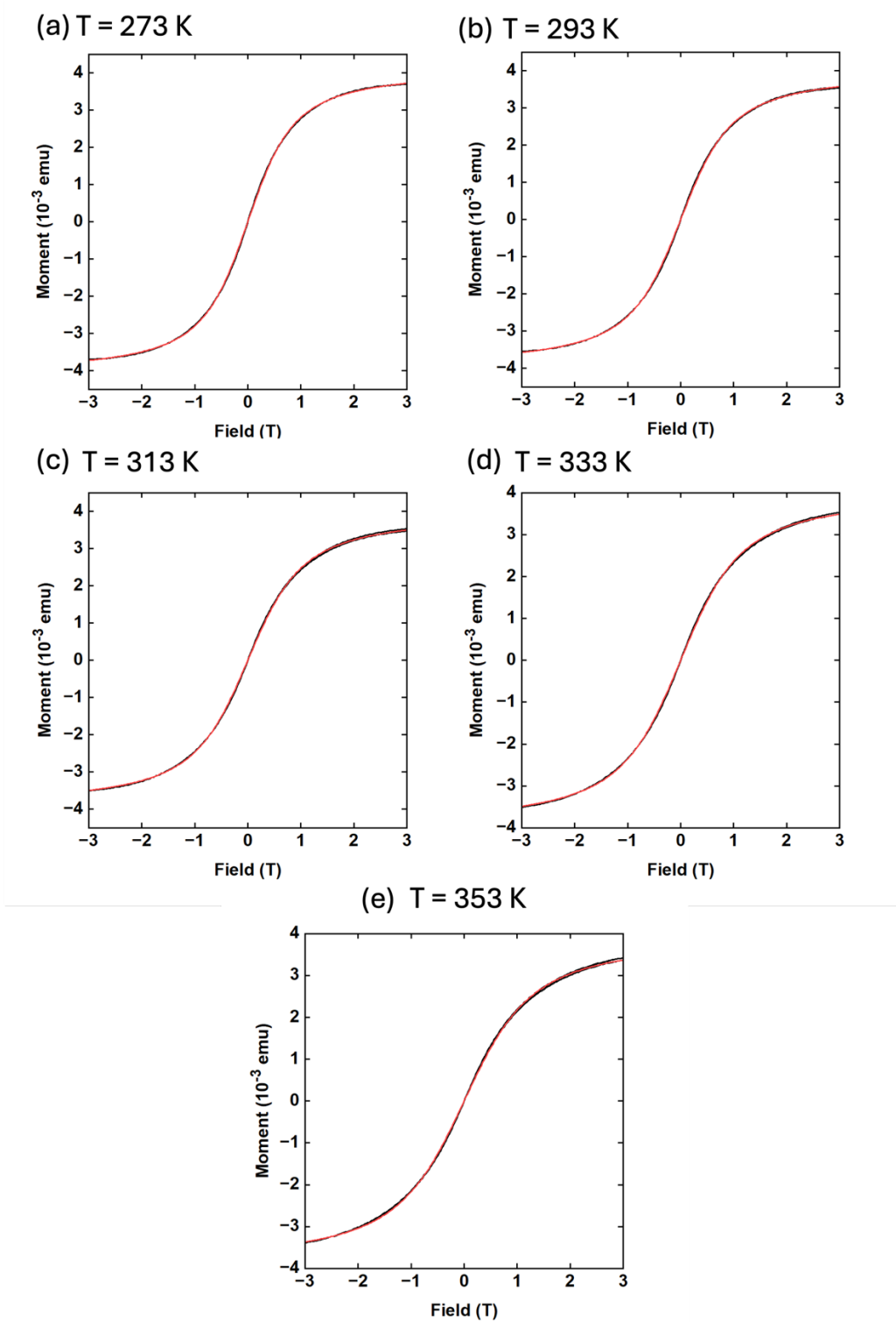

Figure D.2. Data (black line) is fit with modified Langevin function (red line). The best-fit value of $\chi$ is $1.5 \times 10^{-8}$ J/T$^2$ and that of $N$ is $3.88 \times 10^{14}$ at 273 K and 293 K; $3.89 \times 10^{14}$ at 313 K; $4.11 \times 10^{14}$ at 333 K; $4.27 \times 10^{14}$ at 353 K. The 95 % CI stands for 95 % confidence interval of the fitting.

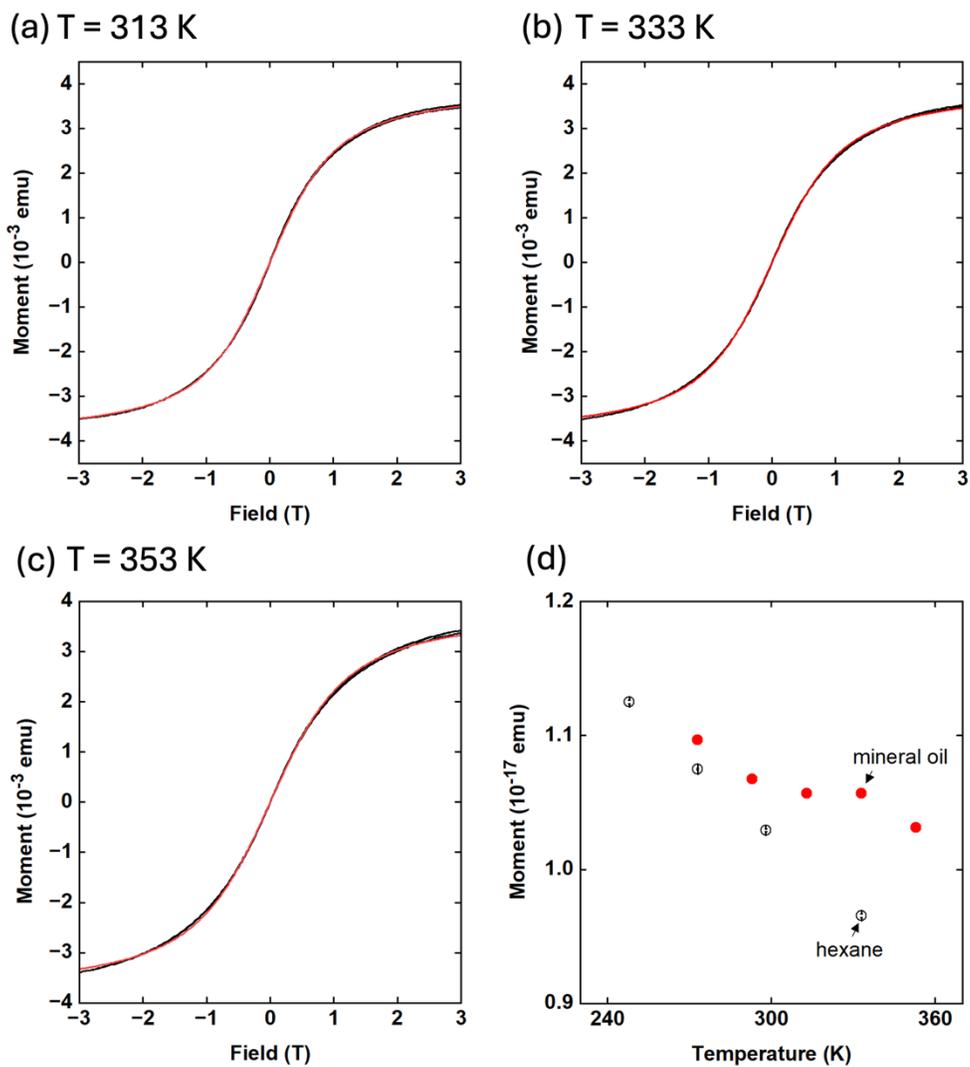

Figure D.3. (a)-(c) Data (black line) is fit with modified Langevin function (red line). $N$ is fixed as $3.88 \times 10^{14}$ at 313 K, 333 K, and 353 K, same as 273 K and 293 K. (d) The temperature dependence of magnetic moment in mineral oil by fixing N for fitting at all temperatures.